\documentclass[journal]{IEEEtran}
\usepackage{cite}
\usepackage{amsmath,amssymb,amsfonts}
\usepackage{algorithmic}
\usepackage{graphicx}
\usepackage{caption}
\usepackage{multirow}
\usepackage{textcomp}
\usepackage{marginnote}
\usepackage{booktabs} 
\usepackage{color,xcolor}
\begin{document}
%
\title{A Survey on Sampling and Profiling over Big Data (Technical Report)}
%
%
%
\author{{Zhicheng Liu}$^{*}$,
	{Aoqian Zhang}$^{\$}$\\
$^{*}$Tsinghua University,  lzc18@mails.tsinghua.edu.cn\\
$^{\$}$University of Waterloo, aoqian.zhang@uwaterloo.ca	}
\maketitle

\begin{abstract}
Due to the development of internet technology and computer science, data is exploding at an exponential rate. Big data brings us new opportunities and challenges. On the one hand, we can analyze and mine big data to discover hidden information and get more potential value. On the other hand, the 5V characteristic of big data, especially Volume which means large amount of data, brings challenges to storage and processing. For some traditional data mining algorithms, machine learning algorithms and data profiling tasks, it is very difficult to handle such a large amount of data. The large amount of data is highly demanding hardware resources and time consuming. 
Sampling methods can effectively reduce the amount of data and help speed up data processing. Hence, sampling technology has been widely studied  and used in big data context, e.g., methods for determining sample size, combining sampling with big data processing frameworks. Data profiling is the activity that finds metadata of data set and has many use cases, e.g., performing data profiling tasks on relational data, graph data, and time series data for anomaly detection and data repair. However, data profiling is computationally expensive, especially for large data sets. Therefore, this paper focuses on researching sampling and profiling in big data context and investigates the application of sampling in different categories of data profiling tasks. 
From the experimental results of these studies, the results got from the sampled data are close to or even exceed the results of the full amount of data. Therefore, sampling technology plays an important role in the era of big data, and we also have reason to believe that sampling technology will become an indispensable step in big data processing in the future.
\end{abstract}


%
\IEEEpeerreviewmaketitle

\section{Introduction}
\label{sec:introduction}
\IEEEPARstart{W}{e} are in the era of big data. With the development of computer science and internet technology, data is exploding at an exponential rate. According to statistics, Google processes more than hundreds of PB data per day, Facebook generates more than 10 PB of log data per month, Baidu processes nearly 100 PB of data per day, and Taobao generates dozens of terabytes online transaction data every day \cite{DBLP:journals/monet/ChenML14}. In May 2011, the McKinsey Global Institution (MGI) released the report\footnote{The Next Frontier of Big Data: Innovation, Competition, and Productivity} which said that big data has great potential in the European Public Sector, US Health Care, Manufacturing, US Retail Industry and Location-based Services. MGI estimates in the report that the mining and analysis of big data will generate 300 billion in potential value per year in the US medical sector and more than 149 billion in the European public sector \cite{manyika2011big}. It can be seen that there is great value behind big data. Therefore, mining the hidden value under big data makes a lot of sense.

Big data is something so huge and complex that it is difficult or impossible for traditional systems and tools to process and work on it \cite{anuradha2015brief}. In the latest development, IBM uses "5Vs" model to depict big data. In the "5Vs" model, Volume means the amount of data and it is the most direct difficulty faced by traditional systems; Velocity means that data is generated quickly; Variety means that data sources and data types are diverse including structural, semi-structured, and unstructured data; Value is the most important feature of big data, although the value density of data is low; Veracity refers to that data quality of big data where there is dirty data. Because big data is so large that data analysis and data mining based on big data require high computing power and storage capacity. In addition, some classical mining algorithms that require several passes over the whole dataset may take hours or even days to get result \cite{DBLP:conf/kdd/ChenHS02}. 

\subsection{Data Sampling}

At present, there are two major strategies for data mining and data analysis: sampling and using distributed systems \cite{DBLP:journals/informaticaSI/Bifet13}. The existing big data processing framework includes batch processing framework like Apache Hadoop, streaming data processing framework like Apache Storm, hybrid processing framework like Apache Spark and Apache Flink. Sampling is a scientific method of selecting representative sample data from target data. Designing a big data sampling mechanism is to reduce the amount of data to a manageable size for processing \cite{DBLP:journals/tkde/WuZW014}. Even if computer clusters are available, we can use sampling such as block-level sampling to speed up big data analysis \cite{9007871}.

Different from distributed systems, sampling is a kind of data reduction method like filtering. Distributed systems increase computing power by adding hardware resources. 
However, a huge computing cost is not always affordable in practice. It is highly demanded to perform the computing under limited resources. 
In this sense, 
sampling is very useful. Since the full amount of data is not used, the approximate result is obtained from the sample data. Such approximate result is quite useful in the context of big data. The computational challenge of big data means that sampling is essential and the sampling methods chosen by researchers is also important \cite{gonzalez2013social}. Besides, the biases caused by sampling are also something need to be considered. 

Sampling or re-sampling is to use less data to get the overall characteristics of the whole dataset. Albattah \cite{DBLP:conf/bdaw/Albattah16} studies the role of sampling in big data analysis. He believes that even if we can handle the full amount of data, we don't have to do this. They focus on how sampling will play its role in specific fields of Artificial Intelligence and verify it by doing experiments. The experimental results show that sampling not only reduces the data processing time, but also get better results in some cases. Even though some examples of sampling are not as effective as the original dataset, they are obviously negligible compared to the greatly reduced processing time. As stated in \cite{DBLP:conf/bdaw/Albattah16}, we believe that sampling can improve big data analysis and will become a preprocessing step in big data processing in the future. 

When it comes to sampling, how to determine the sample size is a very important factor, and different scholars have proposed many methods to determine the sample size \cite{singh2014sampling, DBLP:conf/ccece/Satyanarayana14, DBLP:conf/kdd/JohnL96, Singh2017A}. 
And we also have to consider sampling bias when using sampling techniques.  In addition, some scholars have also studied the application of sampling techniques in the context of big data, e.g., combining sampling with distributed storage, big data computing frameworks. And these will be introduced in detail in Section \ref{Sampling-Techniques}.

\subsection{Data Profiling}

Data mining is an emerging research area, whose goal is to extract significant patterns or interesting rules from large data sets \cite{DBLP:conf/ride/ZakiPLO97}. Data profiling gathers metadata of data that can be used to find data to be mined and import data into various tools for analysis, which is an important preparatory task \cite{DBLP:books/mk/Pyle99}. There is currently no formal, universal or widely accepted definition of distinction between data profiling and data mining. 
Abedjan et al. \cite{DBLP:journals/vldb/AbedjanGN15} think data profiling is used to generate metadata for data sets that are used to help understand data sets and manage data sets. However, data mining is used to mine the hidden knowledge behind the data, which is not so obvious. Of course, data profiling and data mining also have some overlapping tasks, such as association rule mining and clustering. In summary, the goal of data profiling is to generate summary information about the data to help understand the data, and the goal of data mining is to mine the new insights of the data.

There are many use cases of data profiling, 
such as 
data profiling for 
missing data imputation \cite{DBLP:journals/pvldb/SongZC015,DBLP:journals/tkde/SongSZCW20}
or 
erroneous data repairing in relational database \cite{DBLP:conf/sigmod/SongZW16}.
However, data profiling itself has to face computational challenges, especially when it comes to large data sets. Hence, how to alleviate the computational challenges of data profiling is very significant in era of big data. As mentioned above, sampling for big data profiling is very valuable and meaningful. 
We will give a brief introduction for data profiling in Section \ref{data-profiling}.

\subsection{Sampling for Data Profiling}
In this paper, we focus on the sampling techniques used for big data profiling. Certainly, we will first introduce data profiling and sampling technology separately. Among them, data profiling has been associated with outstanding survey papers such as \cite{DBLP:journals/vldb/AbedjanGN15}.  Finally, our core content is to introduce the application of sampling in data profiling tasks when facing large data sets.

\begin{figure}[!t]
	\centering
	\includegraphics[width=3.4in]{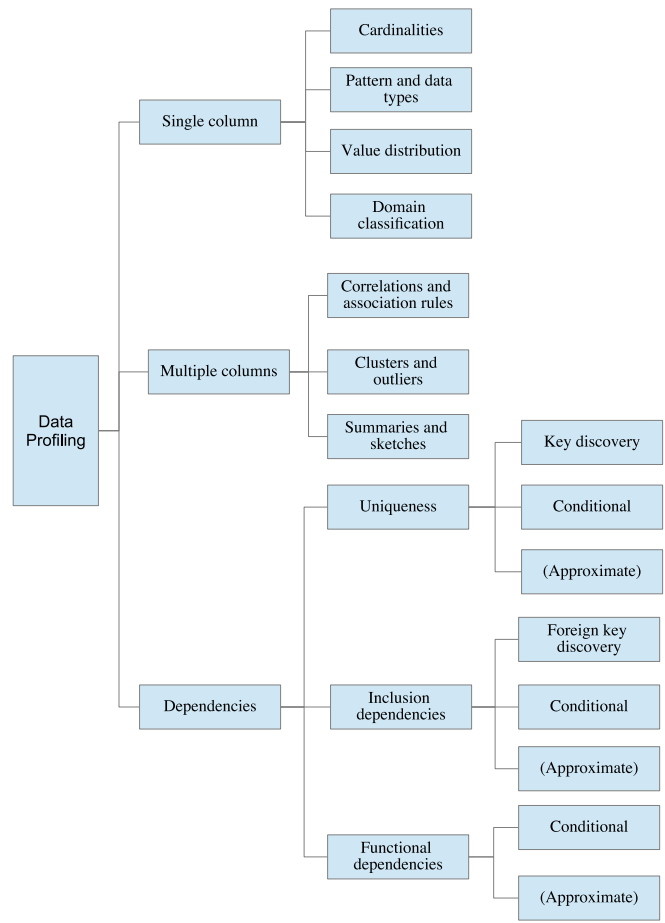}
	\caption{A classification of typical data profiling tasks \cite{DBLP:journals/vldb/AbedjanGN15}.\label{fig:data_profiling}}
\end{figure}

In \cite{DBLP:journals/vldb/AbedjanGN15}, the research on data profiling around the relational database is fully investigated and introduced. The classification of data profiling (see Figure \ref{fig:data_profiling}) is given in \cite{DBLP:journals/vldb/AbedjanGN15}. We will investigate the sampling techniques for important data profiling tasks in single column, multiple columns and dependency according to the classification of data profiling in \cite{DBLP:journals/vldb/AbedjanGN15}. Some traditional sampling methods are introduced in \cite{singh2014sampling}, and methods of determining the sample size are mainly introduced, but less attention is paid to sampling in big data context. Therefore, when discussing the sampling technology below, we will supplement some applications and information of sampling in the big data scenario, e.g., block-based sampling.

\begin{figure}[!t]
	\centering
	\includegraphics[width=3.4in]{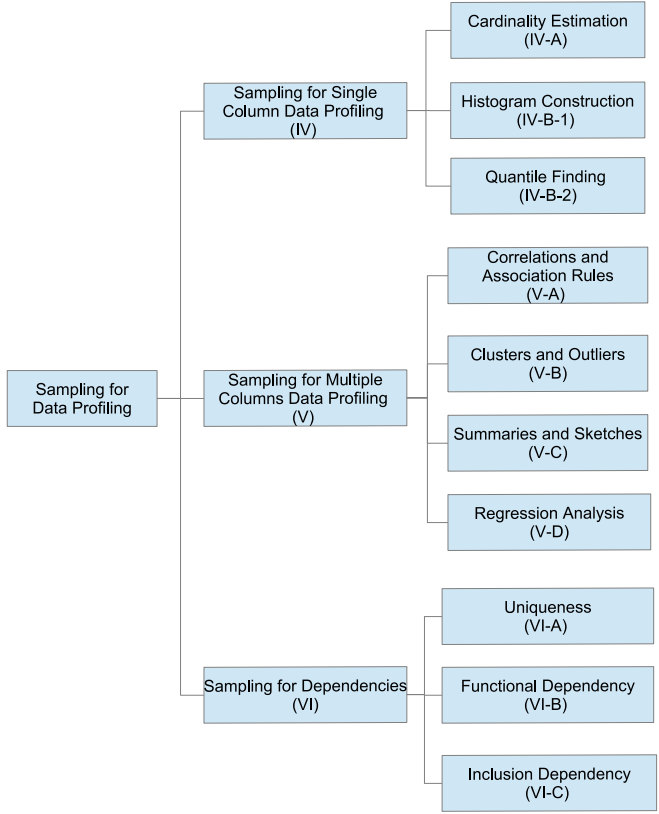}
	\caption{A systematic search method for selecting studies, following the categorization in Figure \ref{fig:data_profiling} by \cite{DBLP:journals/vldb/AbedjanGN15}.}
	\label{fig:structure}
\end{figure}

Specifically, in order to ensure the comprehensiveness of the survey, we follow the systematic search method provided in \cite{DBLP:journals/vldb/AbedjanGN15}, a comprehensive summary of data profiling techniques. As also illustrated in Figure \ref{fig:data_profiling} of our manuscript, Abedjan et al. \cite{DBLP:journals/vldb/AbedjanGN15} categorize the data profiling approaches into three aspects, from the elementary columns to the complex ones, i.e., (1) data profiling for single columns, (2) data profiling for multiple columns, and (3) data profiling for dependencies. While the sampling techniques for data profiling are not emphasized in \cite{DBLP:journals/vldb/AbedjanGN15}, in our paper, we extensively select the studies on sampling for data profiling in the aforesaid categories, respectively. Figure \ref{fig:structure} presents the systematic search method for selecting studies, following the categorization in [13]. Following this method, we summarize the typical methods selected in each category in Table \ref{table:summary-of-sampling}.

The remaining of this paper is organized as follows. 
In Section \ref{data-profiling}, we introduce  
the relevant knowledge of data profiling. In Section \ref{Sampling-Techniques}, we introduce sampling techniques and some important factors in sampling techniques.
Next we introduce the application of sampling for single-column data profiling tasks in Section \ref{single-column}, multi-column data profiling tasks in Section \ref{multiple-columns} and dependencies in Section \ref{dependencies} based on the classification of data profiling tasks in \cite{DBLP:journals/vldb/AbedjanGN15}. Finally, in Section \ref{summary}, we summarize the content of the article and propose some future works. The organizational structure of this article is shown in Figure \ref{fig:structure}.

\section{Data Profiling}
\label{data-profiling}
Before using or processing data, it is very important to have a general understanding of the data. Data profiling is the activity that finds metadata of data set \cite{DBLP:journals/sigmod/Naumann13,DBLP:journals/vldb/AbedjanGN15, DBLP:conf/icde/AbedjanGN16}, 
therefore it can provide basic information about data to help people understand the data. Data profiling is an important area of research for many IT experts and scholars. Data profiling has many classic use cases, such as data integration, data quality, data cleansing, big data analysis, database management, query optimization \cite{DBLP:journals/sigmod/Naumann13, DBLP:journals/vldb/AbedjanGN15}. Abedjan et al. \cite{DBLP:journals/vldb/AbedjanGN15} mainly investigates data profiling for relational data. However, in addition to relational databases, many non-relational databases need data profiling \cite{DBLP:journals/sigmod/Naumann13}, 
such as
time series data \cite{DBLP:conf/sigmod/SongZWY15,DBLP:conf/sigmod/ZhangSW16,DBLP:journals/access/WangLML20},
graph data \cite{DBLP:conf/sigmod/FanWX16a,DBLP:journals/pvldb/SongCY014,DBLP:journals/vldb/SongLCYC17}, 
or 
heterogeneous data in dataspaces \cite{DBLP:conf/kr/MaierHF06,DBLP:conf/icde/SongCY11,DBLP:journals/vldb/Song0Y13}.

Data profiling tasks are classified in \cite{DBLP:journals/vldb/AbedjanGN15} and \cite{DBLP:journals/sigmod/Naumann13}. Abedjan et al. \cite{DBLP:journals/vldb/AbedjanGN15} classify the data profiling tasks of single data source, and divides the tasks of data profiling into single column data profiling, multiple columns data profiling and dependency (see Figure \ref{fig:data_profiling}). In fact, dependencies belong to multiple columns data profiling tasks. Abedjan et al. \cite{DBLP:journals/vldb/AbedjanGN15} put dependencies separately into a large category and discuss it in detail. Naumann \cite{DBLP:journals/sigmod/Naumann13} classifies data profiling from single data source to multiple data sources. 

There are three challenges for data profiling: managing the input, performing the computation and managing the output \cite{DBLP:journals/sigmod/Naumann13, DBLP:journals/vldb/AbedjanGN15, DBLP:conf/sigmod/AbedjanGN17}. 
In this article we focus on the second challenge, performing the computation, i.e., the computational complexity of data profiling. The computational complexity of data profiling depends on the number of rows and columns of data. When the data set is very large, the calculation of data profiling can be very expensive. This is why we care about sampling for big data profiling, in order to reduce the computational pressure and speed up the process of data profiling.

\section{Sampling Techniques}
\label{Sampling-Techniques}



In this section, we introduce common sampling techniques in \ref{common-sampling}, application of sampling technology in big data context in \ref{sampling-in-big-data-context}, methods of determining sample size in \ref{determination-of-sample-size} and resolutions of reducing sampling bias in \ref{resolution-for-sampling-bias}.

\subsection{Common Sampling Techniques}
\label{common-sampling}
Sampling refers to estimating the characteristics of the entire population through the representative subsets within the population \cite{singh2014sampling}. From a big perspective, sampling involves probability and non-probability sampling. Probability sampling means that every unit in a finite population has a certain probability to be selected, and it does not necessarily require equality. Non-probability sampling is generally based on subjective ideas and inferences, e.g., common web questionnaires \cite{schreuder2001applications, iachan2019weighting}. The sampling methods mentioned below are all probability sampling methods. Sampling is often used in data profiling \cite{DBLP:journals/vldb/AbedjanGN15}, data analysis \cite{DBLP:journals/spm/SlavakisGM14}, data mining \cite{DBLP:journals/tkde/WuZW014}, data visualization \cite{DBLP:conf/medes/AgrawalKDA15}, machine learning \cite{DBLP:journals/ijon/ZhouPWV17} etc. The advantage of sampling is that algorithms or models can be conducted using subset instead of the whole data set. There are some commonly used sampling techniques including simple random sampling \cite{DBLP:journals/amc/KadilarC04}, stratified sampling \cite{bickel1984asymptotic}, systematic sampling \cite{gundersen1987efficiency}, cluster sampling \cite{henderson1982cluster}, oversampling and undersampling \cite{DBLP:conf/kdd/LeskovecF06, DBLP:conf/daeng/YapRRFKA13}, reservoir sampling \cite{DBLP:journals/toms/Vitter85}, etc. Table \ref{tab:sampling} gives an overview of these common sampling methods.

\begin{table*}[]
	\centering
	\caption{Common sampling methods}
	\label{tab:sampling}
	\begin{tabular}{ll}
		\toprule
		Sampling method & Description \\
		\midrule 
		Simple random sampling & Extracting a certain number of samples and each tuple is selected with equal probability.   \\
		
		Stratified random sampling     & Tuples are divided into homogeneous groups and sample from each group.   \\
		
		Systematic sampling            & Sampling at regular intervals until the sample size is satisfied.                \\ 
		
		Cluster sampling               & Tuples are divided into non-overlapping groups and randomly select some groups as samples.       \\
		
		Oversampling and Undersampling &    Oversampling randomly repeat the minority class samples, while Undersampling randomly discard \\ 
		& the majority class samples to balance the data. \\
		Reservoir sampling             & Adding tuples into the reservoir of a fixed size with unknown size of the entire data set.                             \\
		\bottomrule                                        
	\end{tabular}
\end{table*}


\subsection{Sampling in Big Data Context}
\label{sampling-in-big-data-context}
\begin{figure}[!t]
	\centering
	\includegraphics[width=3.4in]{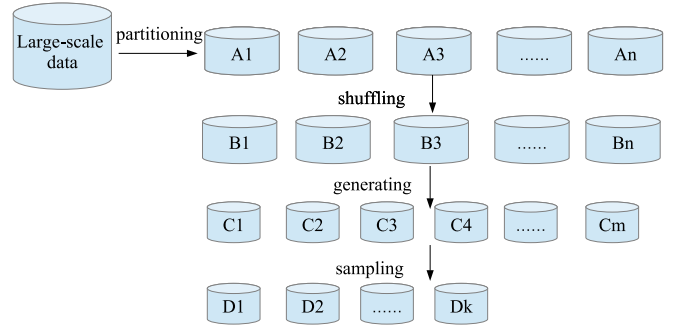}
	\caption{I-sampling workflow \cite{DBLP:conf/bigdata/HeHLWW17}.}
	\label{block-based_sampling.png}
\end{figure}

In the era of big data, the application of sampling is particularly important due to the large amount of big data. And sampling can be performed with the help of big data computing framework. For example, He et al. \cite{DBLP:journals/fss/HeWZSS15} use MapReduce to sample from the data which contains uncertainty. He et al. \cite{DBLP:conf/bigdata/HeHLWW17} propose a block-based sampling (I-sampling) method for large-scale dataset when the whole dataset is already assigned on a distributed system. The processing flow of I-sampling is shown in Figure \ref{block-based_sampling.png}.


The traditional sampling methods like simple random sampling, stratified sampling and systematic sampling are records-based. These records-based sampling methods require the complete pass over the whole dataset. Hence, they are commonly used for small or medium scale datasets on single computer. Even though the whole dataset is already assigned on a distributed system, it is very difficult to get a high-quality partition of the original dataset \cite{DBLP:journals/cgf/LiuJH13}. In the era of big data, data profiling tasks can be carried out on distributed systems, e.g., data profiling tasks on HDFS data. Therefore, block-based sampling proposed in \cite{DBLP:conf/bigdata/HeHLWW17} can be used as a promising sampling method for data stored in distributed machines.  

He et al. \cite{DBLP:conf/bigdata/HeHLWW17} propose a block-based sampling method for large-scale dataset. They take block-based sampling as one of components of their big data learning framework which is called the asymptotic ensemble learning framework \cite{DBLP:conf/bdc/SalloumHH16}. However, the block-based sampling method is suitable for data that is randomly ordered and not for those records that are stored in an orderly manner. In order to solve this problem, they propose a general block-based sampling method named I-sampling. 

I-sampling contains four steps to get sample. Firstly, I-sampling divides large-scale dataset into non-overlapping primary data blocks A$_{i}$. Secondly, I-sampling shuffles primary data blocks A$_{i}$ to get shuffling data blocks B$_{i}$. The purpose of shuffling is to disrupt the order of original data. Thirdly, I-sampling randomly selects data from B$_{i}$ and put it into the basic blocks to get a block pool $C$. Finally, a certain number of basic blocks are randomly selected from the block pool, and the data in these blocks is taken as sample. In experiments, He et al. \cite{DBLP:conf/bigdata/HeHLWW17} demonstrate that the block-based sampling has the basically equal means and variances with simple random sampling. Besides, the distribution of I-sampling data is approximately the same with original dataset. And RMSEs of extreme learning machine based on records-based random sampling and I-sampling are nearly the same. The processing flow of I-sampling is shown in Figure \ref{block-based_sampling.png}.

As a matter of fact, data contains uncertainty in many applications. For example, when we do experiments, such as sampling, uncertainty occurs because there are many potential results for sampling. Uncertainty means the diversity of potential outcomes, which is unknown to us. And dealing with big data with uncertainty distribution is one of the most important issues of big data research \cite{DBLP:journals/fss/HeWZSS15}. The sample quality affects the accuracy of data profiling. The following example shows how to use MapReduce to accelerate sampling from a big data set with uncertainty distribution, and select Minimal Consistent Subset with better sample quality. 
Minimal Consistent Subset (MCS) is a consistent subset with a minimum number of elements.

He et al. \cite{DBLP:journals/fss/HeWZSS15} use MapReduce to sample from the data which contains uncertainty. They propose a Parallel Sampling method based on Hyper Surface (PSHS) for big data with uncertainty distribution to get the MCS of the original sample set. Hyper Surface Classification (HSC) is a classification method based on Jordan Curve Theorem and put forward by He et al. \cite{he2003novel}. The MCS of HSC is a sample subset by selecting one and only one representative sample from each unit included in the hyper surface. Some samples in the MCS are replaceable, while others are not, leading to the uncertainty of elements in MCS \cite{DBLP:journals/fss/HeWZSS15}. Because of large-scale of data, they use MapReduce for parallel sampling. MapReduce is a well-known distributed computing framework for big data today.

PSHS algorithm needs to execute three kinds of MapReduce jobs. In the first task, based on the value of each dimension of the data, the map function places each sample in the region to which it belongs. The reduce function determines whether this area is pure, and labels each area with a corresponding label: pure or impure. In the second task, the corresponding decision tree is generated and the samples that have no effect on the generated decision tree must be removed. The third task is the sampling task, where the map function is to place each sample in the pure region it belongs to according to the rules. In pure regions, these samples have the same effect on building the classifier, hence the reduce task is to randomly select one and only one from each region for building the MCS. The Minimal Consistent Subset selected by this parallel sampling is a good representation of the original data.

\subsection{Determination of Sample Size}
\label{determination-of-sample-size}
It is very important to select effective samples \cite{DBLP:journals/jbd/TsaiLCV15}. If the sample size is too small, it may get an incorrect conclusion. If the sample size is too large, the calculation time is too long. 
Therefore, when performing machine learning algorithms, data mining algorithms or data profiling tasks on large-scale dataset, how to choose the appropriate sampling method and determine the sample size are important factors in determining whether the correct result (within the allowable error range) can be obtained.  

%
There are some classic methods to determine the sample size. Singh and Masuku \cite{singh2014sampling} have detailed and summarized these traditional methods. For example, you can take the sample size in other similar studies as the size of the sample in your study. 
Furthermore, you can determine the size of the sample according to the published tables. These tables determine the size of sample according to some given evaluation indicators and the size of the original dataset. Some commonly used evaluation indicators include the level of precision, the level of confidence or risk, the degree of variability, etc. 
Another method to assure size of sample is to calculate the size of sample according to some simple calculation formulas, which calculate the size of sample based on sampling error, confidence, and P-value. A simple formula (\ref{calSampleSize}) for estimating the sample size \cite{singh2014sampling} is as follows, where n is the sample size, $N$ is the amount of raw data, $e$ is the level of precision, a 95\% confidence level and P = .5 are assumed.:
\begin{equation}
\label{calSampleSize}
n=\frac{N}{1+N*e^2}
\end{equation}

When data mining algorithms are performed based on massive amounts of data, much of current research prefers to scale up data mining algorithms to deal with computational (time and memory) constraints, but some scholars focus on selecting how many samples to conduct data mining algorithms. 
In data mining algorithms, a common formula used to estimate the number of samples is Probably Close Enough (PCE). The convergence conditions are determined using PCE standard to obtain the best sample size for sampling. PCE is calculated as formula (\ref{cal-sample-size}).
\begin{equation}
\label{cal-sample-size}
\mathit{Pr}[|\mathit{acc(D)}-\mathit{acc(D_i)}|\geq \epsilon] \leq \delta
\end{equation}
Where D$_i$ represents sample data, D represents all data, $\epsilon$ represents the error range threshold of accuracy, and $\delta$ represents probability. 

Furthermore, Satyanarayana \cite{DBLP:conf/ccece/Satyanarayana14} proposes a dynamic adaptive sampling method for estimating the amount of data required for the learning curve to converge at each iteration. The author applies Chebyshev inequality to derive an expression that will estimate the number of instances at each iteration, which takes advantage of classification accuracy in order to get more precise estimates of the next sample. The expression is formula (\ref{Chebyshev's inequality}), where D$_i$ is sample under consideration, acc(x$_i$) is classification accuracy of each instance, $\epsilon$ is approximation parameter and $\delta$ is probability of failure. And Satyanarayana \cite{DBLP:conf/ccece/Satyanarayana14} uses the formula (\ref{convergence-equation}) to check convergence at each iteration, where D$_i$ is the sample of current iteration and D$_{i-1}$ is the sample of last iteration.
\begin{equation}
\label{Chebyshev's inequality}
m \geq \frac{2}{\frac{1}{|D_i|}\sum_{i=1}^{D_i}acc(x_i)}[\frac{1}{\epsilon^2}\log{\frac{1}{\delta}}]
\end{equation}
\begin{equation}
\label{convergence-equation}
|\frac{1}{|D_i|}\sum_{i=1}^{D_i}acc(x_i)-\frac{1}{|D_{i-1}|}\sum_{i=1}^{D_{i-1}}acc(x_i)| \textless \epsilon
\end{equation}

\begin{figure}[!t]
	\centering
	\includegraphics[width=3.4in]{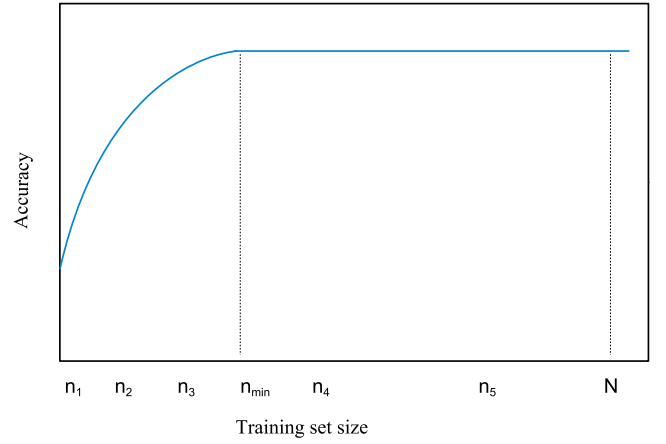}
	\caption{Learning curves \cite{DBLP:conf/kdd/ProvostJO99}.}
	\label{progressive_sampling.png}
\end{figure}

When sampling is used in machine learning, the most appropriate number of samples is to make the accuracy rate reach the maximum value and increasing the number of samples can no longer improve the accuracy of the learning algorithm. The corresponding figure is Figure \ref{progressive_sampling.png}, where n$_{min}$ is the minimum sample size. In this case, 
there is a method for determining the minimum number of samples called sequential sampling.
Sequential sampling refers to sample sequentially and stop sampling until a certain criterion is met. John and Langley \cite{DBLP:conf/kdd/JohnL96} propose a method called Arithmetic Sampling. This method uses a schedule Sa = (n$_{0}$, n$_{0}$ + n$_{\delta}$, n$_{0}$ + 2n$_{\delta}$, n$_{0}$ + 3n$_{\delta}$, ……, N) to find the minimum sample size, where  n$_{0}$ is the starting sample size and n$_{\delta}$ is fixed interval. Provost et al. \cite{DBLP:conf/kdd/ProvostJO99} think that the main drawback is that if the minimum size of sample is a large multiple of n$_{\delta}$, it will take many runs to reach convergence. 
Obviously, if n$_{\delta}$ is too small, it will take many iterations to get convergence, and if n$_{\delta}$is too large, it may skip the optimal size of sample.

Singh et al. \cite{Singh2017A} propose another sequential sampling strategy for classification problem. They mention that data for training machine learning models typically originates from computer experiments such as simulations. And computer simulations are often computationally expensive. In order to ease the computation pressure, they use sampling to get as little data as possible. 
The sequential sampling starts with an initial small data set X$^{\delta}$, and it will iteratively increase the sample by taking training points at well-chosen locations ${\delta}$ in the input space until stopping criteria is reached. The sequential sampling strategy chooses a representative set of data samples that focuses the sampling on those locations in the input space where the class labels are changing more rapidly, while making sure that no class regions are missed \cite{Singh2017A}. The sample update formula is formula (\ref{sequential-2}) where class labels Y$^{\delta}$ obtained by formula (\ref{sequential-1}) are result of simulator evaluates X$^{\delta}$. With sequential sampling strategy, small and high quality samples can be obtained.
\begin{equation}
\label{sequential-1}
Y^\delta := f(X^\delta)
\end{equation}
\begin{equation}
\label{sequential-2}
S := S \cup (X^\delta, Y^\delta)
\end{equation}

\subsection{Sampling Error and Sampling Bias}
\label{resolution-for-sampling-bias}
Sampling error is when a randomly chosen sample does not reflect the underlying population purely by chance and sampling bias is when the sample is not randomly chosen at all \cite{harford2014big}. Sampling bias is one of the causes of sampling error. These two are often confused by some scholars. Sampling bias is caused by the failure of the sampling design, which cannot truly extract the sample randomly from the population \cite{nagler2015drawing}. 

There is a typical case of sampling error. The large Nurses Health Study tracked 48,470 postmenopausal women for 10 consecutive years, aged between 30 and 63 years old. The study concluded that hormone replacement therapy can reduce the incidence of severe coronary heart disease by nearly half \cite{stampfer1991postmenopausal}. Despite the large sample size, the study failed to recognize the atypical nature of the sample and the confusion of estrogen therapy with other active health habits \cite{kaplan2014big}. This also illustrates the importance of proper sampling methods and the collected samples to get the right conclusions.

To be able to correctly select the samples that represent the original data, 
Kim and Wang \cite{Kim2018Sampling} focus on and solve the problem of selection bias in the process of sampling.  Since big data is susceptible to selection bias, they propose two ways to reduce the selection bias. 
One is based on the inverse sampling method. This method is divided into two stages. The first stage is to sample directly from the big data. The sample is easily affected by the selection bias, thereby it is necessary to calculate the importance of each element in the sample to determine selection bias. In the second stage, the data sampled from the first stage is re-sampled according to the importance weight of each element. In this way, they have achieved the goal of realizing the correction of the selection deviation. The other is the idea of using data integration. They propose to use the survey data and big data to correct the selection bias by means of the auxiliary information of survey data.
%

From the perspective of official statisticians, Tam et al. \cite{tam2018big} believe that big data is challenged by self-selection bias. Self-selection bias causes biased sample with non-probability sampling. Inferences from big data with this bias will be affected. Thus, they outline methods for adjusting self-selection bias to estimate proportions, e.g., using pseudo weights and super population models \cite{MaiselINFERENCE}. 

As a matter of fact, the case of 2016 US presidential election studied in \cite{meng2018statistical} is precisely because of the existence of self-selection bias, which ultimately leads to data deceiving us. Therefore, to get the correct conclusion from the data, you need to ensure the quality of the data. Probability sampling can guarantee the quality of the data. 
When probability sampling cannot be satisfied, the data will be affected by Law of Large Populations (LLP). The large amount of data N will affect our estimation error. 
In summary, when doing data sampling, data quality must be taken into account, and those high quality data sets should be given higher weight. This will prevent our statistical inferences from being affected.

\begin{table*}[]
	\centering
	\caption{Overview of single-column profiling tasks \cite{DBLP:journals/vldb/AbedjanGN15}}
	\label{table:single-column}
	\begin{tabular}{lll}
		\toprule
		Category                                                                                     & Task         & Description                                                                                                                           \\  \midrule 
		Cardinalities                                                            & num-rows     & Number of rows                                                                                                                        \\
		& value length & Measurements of value lengths (minimum, maximum, median, and average)                      \\
		& null values  & Number or percentage of null values                                                                                                   \\
		& distinct     & Number of distinct values; sometimes called "cardinality"                                                                            \\
		& uniqueness   & Number of distinct values divided by the number of rows                                                                              \\
		Value distributions                                                      & histogram    & Frequency histograms (equi-width, equi-depth, etc.)                                                                                   \\
		& constancy    & Frequency of most frequent value divided by number of rows                                 \\
		& quartiles    & Three points that divide the (numeric) values into four equal groups                     \\
		& first digit  & Distribution of first digit in numeric values                                                                                         \\
		Patterns, data types, and domains & basic type   & Generic data type, such as numeric, alphabetic, alphanumeric, date, time        \\
		& data type    & Concrete DBMS-specific data type, such as varchar, timestamp.                              \\
		& size         & Maximum number of digits in numeric values                                                                                            \\
		& decimals     & Maximum number of decimals in numeric values                                                                                          \\
		& patterns     & Histogram of value patterns (Aa9...)                                                                                                  \\
		& data class   & Semantic, generic data type, such as code, indicator, text, date/time, quantity, identifier\\
		& domain       & Classification of semantic domain, such as credit card, first name, city, phenotype       \\ 
		\bottomrule
	\end{tabular}
\end{table*}

\begin{table*}[]
	\centering
	\caption{Summary of sampling for big data profiling tasks}
	\label{table:summary-of-sampling}
	\begin{tabular}{lll}
		\toprule
		\multicolumn{2}{l}{Data Profiling}                                     & Sampling-based method                                                                                               \\ \midrule
		Single column    &  Cardinality Estimation              & \^{D$_{hybrid}$} \cite{DBLP:conf/vldb/HaasNSS95}, GEE \cite{DBLP:conf/pods/CharikarCMN00}, AE \cite{DBLP:conf/pods/CharikarCMN00}, Distinct sampling \cite{DBLP:conf/vldb/Gibbons01}                                                                                               \\
		&  Histograms                          & Random sampling \cite{DBLP:conf/sigmod/ChaudhuriMN98}, Backing sample \cite{DBLP:journals/tods/GibbonsMP02}                                                                                   \\
		& Quantiles                          & Non-uniform random sampling \cite{DBLP:conf/sigmod/MankuRL99}, Improved random sampling \cite{DBLP:conf/sigmod/HuangWYL11}                                                                                 \\
		Multiple columns & Correlations and association rules & Sequential random sampling without replacement \cite{DBLP:conf/ride/ZakiPLO97}, Two-phased sampling \cite{DBLP:conf/kdd/ChenHS02}, ISbFIM \cite{DBLP:journals/mlc/WuFPZY15}\\
		& Clusters and outliers              & Biased sampling \cite{DBLP:journals/tkde/KolliosGKB03}                                                                                              \\
		& Summaries and sketches             &    Error-bounded stratified sampling \cite{DBLP:journals/pvldb/YanCZ14}, \cite{DBLP:conf/ldav/RojasKRD17}                                                                \\
		& Regression analysis                &   IBOSS \cite{wang2019information}, Random sampling without replacement \cite{jun2015divided}                                                         \\
		Dependency      & Uniqueness                         & GORDIAN \cite{DBLP:conf/vldb/SismanisBHR06}, HCA-Gordian \cite{DBLP:conf/cikm/AbedjanN11}                                                                                                 \\
		& Functional dependencies             & AID-FD \cite{DBLP:conf/cikm/BleifussBFRW0PN16}, HYFD \cite{DBLP:conf/sigmod/PapenbrockN16}, CORDS \cite{DBLP:conf/sigmod/IlyasMHBA04}, BRRSC \cite{8935096}                                                                                           \\
		& Inclusion dependencies            & FAIDA \cite{DBLP:conf/btw/0001PDFHZZN17}                                                                            \\  \bottomrule
	\end{tabular}
\end{table*}

\section{Sampling for Single Column Data Profiling} \label{single-column}
Single column data profiling tasks are divided into cardinalities, value distributions, patterns, data types, and domains \cite{DBLP:conf/bigdataconf/HuangCLS019}. 
Table \ref{table:single-column} \cite{DBLP:journals/vldb/AbedjanGN15} lists typical metadata that may result from single-column data profiling. For some single-column data profiling tasks, such as decimals which calculates maximum number of decimals in numeric values, simple sampling methods cannot guarantee reliable results. And for identifying a domain of one column, it is often more difficult and not fully automated \cite{DBLP:series/synthesis/2018Abedjan}. Among them, cardinality, histograms and quantiles are often used for query optimizers, therefore sampling techniques are more commonly used in these tasks. Specifically, in Section \ref{sect:cardinality}, we introduce sampling for cardinality estimation. Section \ref{sect:distribution} presents sampling for value distribution.
More advanced statistics include the probabilistic correlations on text attributes \cite{DBLP:journals/isci/SongZ014}.

\subsection{Sampling for Cardinality Estimation}
\label{sect:cardinality}
Cardinalities or counts of values in a column are the most basic form of metadata \cite{DBLP:journals/vldb/AbedjanGN15}. Cardinalities usually include number of rows, number of null values and number of distinct values, which is the most important type of metadata \cite{DBLP:journals/pvldb/HarmouchN17}. For some tasks, such as number of rows and number of null values, a single pass over a column can get the exact result. However, finding the number of distinct values may require to sort or hash the value of column \cite{DBLP:series/synthesis/2018Abedjan}.
Similarly, when facing large data sets, statistics of the number of distinct values of an attribute have to face the pressure of memory and calculation. Therefore, the estimation of the number of distinct values based on sampling has been studied \cite{DBLP:conf/vldb/HaasNSS95, DBLP:conf/pods/CharikarCMN00, DBLP:conf/vldb/Gibbons01}.

Haas et al. \cite{DBLP:conf/vldb/HaasNSS95} propose several sampling-based estimators to estimate the number of different values of an attribute in a relational database. They use a large number of attribute value distributions from various actual databases to compare these new estimators with those in databases and statistical literature. Their experimental results prove that no estimator is optimal for all attribute value distributions. And from their experimental results, it can be seen that the larger the sampling fraction, the smaller the estimated mean absolute deviation will be. They therefore propose a sampling-based hybrid estimator \^{D$_{hybrid}$} and get the highest precision on average at a given sampling fraction.

Similar to Haas et al., Charikar et al. \cite{DBLP:conf/pods/CharikarCMN00} also obtain a negative result in the experiment that no estimator based on sampling can guarantee small errors on the input data of different distributions, unless a larger sampling fraction is performed on the input data. They therefore propose a new estimator Guaranteed-Error Estimator (GEE), which is provably optimal. Although its error on the input of different distributions is small, it does not make use of the knowledge of different distributions. For example, in the case of low-skew data with a large number of distinct values, GEE performs not very well in practice. They further propose a new heuristic version of GEE called Adaptive Estimator (AE), which avoids the problems encountered by GEE.

Different from the previous research using random sampling, Gibbons \cite{DBLP:conf/vldb/Gibbons01} proposes distinct sampling to accurately estimate the number of distinct values. Distinct sampling can collect distinct samples in a single scan of the data, and the samples can be kept up to date in the state of data deletions and insertions. On a truly confidential data set Call-center, distinct sampling uses only 1\% of the data, and can achieve a relative error of 1\% -10\%, while increasing the speed of report generation by 2-4 orders of magnitude. They compare distinct sampling with GEE, AE in the experiment and prove that in real-world data sets, distinct sampling performs much better than GEE and AE.

It is worth noting that Harmouch and Naumann \cite{DBLP:journals/pvldb/HarmouchN17} conduct an experimental survey on cardinality estimation. In the experiment, they use the GEE \cite{DBLP:conf/pods/CharikarCMN00} as an example of evaluation. They perform experiments on synthetic and real-world data sets. It can be seen from the experimental results that the larger the sampling fraction, the smaller the average estimation relative error. And when GEE wants to reach 1\% relative error, it needs to collect more than 90\% of the data. 
In conclusion, when faced with large data sets, cardinality estimation requires high memory, and sampling can reduce memory consumption, but cannot guarantee reasonable accuracy all input distributions.

\subsection{Sampling for Value Distribution}
\label{sect:distribution}
Value distribution is a very important part of single-column data profiling. Histogram and quantile are two typical forms used to represent value distribution. The histogram is used to describe the distribution of data, while quantile refers to dividing the data into several equal parts. 

\subsubsection{Sampling for Histogram Construction}
Many commercial database systems maintain histograms to summarize the contents of large relations and permit efficient estimation of query result sizes for use in query optimizers \cite{DBLP:journals/tods/GibbonsMP02}.  Histogram can be used to describe the frequency distribution of attributes of interest, which groups attributes values into buckets and approximates true attribute values and their frequencies based on summary statistics maintained in each bucket \cite{DBLP:phd/Kooi80}. However, the database is updated frequently, hence the histogram also needs to be updated accordingly. Recalculating histograms is expensive and unwise for large relations. 

Gibbons et al. \cite{DBLP:journals/tods/GibbonsMP02} propose sampling-based approaches for incremental maintenance of approximate histograms. They use a "backing sample" to update histograms. Backing sample is a random sample of the relation which is kept up to date in the presence of databases updates, which is generated by uniform random sampling. Therefore, random sampling can help to speed up histogram re-computation. For example, SQL Server recomputes histograms based on a random sample from relations \cite{DBLP:conf/sigmod/ChaudhuriMN98}. 

Chaudhuri et al. \cite{DBLP:conf/sigmod/ChaudhuriMN98} focus on how much sample is enough to construct a histogram. They propose a new error metric called the max error metric for approximate equip-depth histogram. The max error metric is formula (\ref{max_error_metric}) shown below, where b$_j$ is number of values in bucket j, k is the number of buckets and n is the number of records. A k-histogram is said to be a ${\delta}$-deviant histogram when ${\Delta}$max ${\leq}$ ${\delta}$. And size of sample r is calculated as the following formula (\ref{histogram-size}), where ${\delta}$ ${\leq}$ ${\frac{n}{k}}$ and ${\gamma}$ is predefined probability.
\begin{equation}
\label{max_error_metric}
\Delta max =\max \limits_{1\leq j \leq k}|b_j - \frac{n}{k}|
\end{equation}
\begin{equation}
\label{histogram-size}
r \geq \frac{4n^2\ln{(\frac{2n}{\gamma})}}{k\delta^2}
\end{equation}

As mentioned above, the histogram can be used to represent the distribution of data. In exploratory data analysis, analysts want to find a specific distribution from a large number of candidate histograms. The traditional approach is "generate and test", i.e., generating all possible histograms, and then testing whether these histograms meet the requirements. This approach is undesirable when the data set is large. Therefore, Macke et al. \cite{DBLP:journals/pvldb/MackeZHP18} propose a sampling-based approach to identify the top k closest histograms called HistSim. The idea of HistSim is using random sampling method without replacement to collect samples for histogram constructing. Then they normalize the representation vector of the histogram, and use \textit{l1} distance to calculate the similarity.
Furthermore, they propose FastMatch, which combines HistSim and block-based sampling method,  and obtain near-perfect accuracy with up to 35 × speedup over approaches that do not use sampling on several real-world datasets in the experiment.

\subsubsection{Sampling for Quantile Finding}
Quantiles can be used to represent the distribution of single column value. Quantiles are used by query optimizers to provide selectivity estimates for simple predicates on table values \cite{DBLP:conf/sigmod/SelingerACLP79}. Calculating exact quantiles on large data sets is time consuming and requires a lot of memory. For example, quantile finding algorithm in \cite{pohl1969minimum} requires to store at least N/2 data elements to find the median, which is memory unacceptable for large-scale data. 

\begin{figure}[!t]
	\centering
	\includegraphics[width=3.4in]{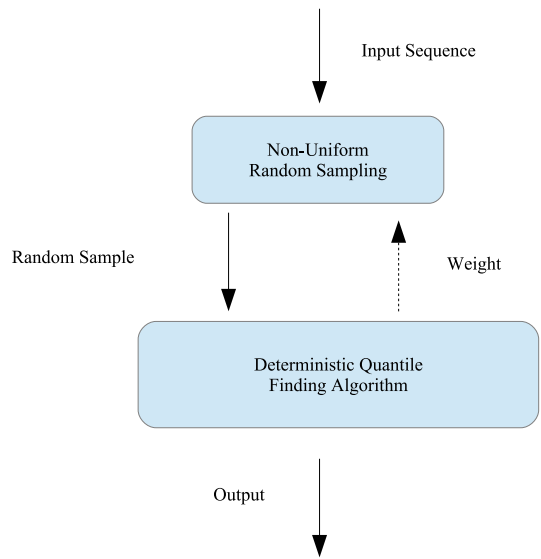}
	\caption{Sampling for quantile finding \cite{DBLP:conf/sigmod/MankuRL99}.}
	\label{Quantile.png}
\end{figure}

Therefore, Manku et al. \cite{DBLP:conf/sigmod/MankuRL99} present a novel non-uniform random sampling to find approximate quantile. They apply non-uniform random sampling to reduce memory requirements. Non-uniform means that the probability of selecting each element in the input is different. They set the earlier elements in the input sequence with larger probability than those arrive later. And the process of quantile finding is shown in Figure \ref{Quantile.png}. When the data arrives, they randomly select an element in each data block and put it into buffers. Then based on sample, deterministic algorithms are performed to find quantiles.

However, simply using random sampling method and calculating the quantiles on the sample may not be accurate enough on sensor networks. Hence, Huang et al. \cite{DBLP:conf/sigmod/HuangWYL11} propose a new sampling-based quantile computation algorithm for sensor networks to reduce the communication cost. To improve accuracy, they augment the random sample with additional information about the data. They analyze how to add additional information to the random sample under the flat model and the tree model. For example, in the flat model, each node first samples each data value independently with a certain probability p and computes its local rank. Then the samples and their local ranks are sent to base station. The base station estimates rank for any value it receives and then quantile queries can be solved. In the end, they prove through experiments that the quantile computation in Sensor Networks based on this new sampling method reduces one to two orders of magnitude in terms of the total communication cost compared with the previous method.

\section{Sampling for Multiple Columns Data Profiling} \label{multiple-columns}
As shown in Figure \ref{fig:data_profiling}, the content of the multiple columns data profiling tasks includes 
association rule mining \cite{DBLP:journals/access/ZhangWZ19}, 
clusters and outliers \cite{DBLP:conf/kdd/SongLZ15}, 
summaries and sketches \cite{DBLP:journals/vldb/AbedjanGN15}. 
Besides, statistical methods such as regression analysis \cite{DBLP:conf/icde/ZhangSSW19} can be used to perform multiple columns analysis, analyzing the relationship between these columns.
Specifically, in Section \ref{sect:association}, we investigate sampling for discovering association rules. Section \ref{sect:cluster} presents the content of sampling for clusters and outliers. And sampling for summaries and sketches is introduced in Section \ref{sect:summaries}. Then, in Section \ref{sect:regression}, we introduce sampling for helping perform regression analysis.

\subsection{Sampling for Discovering Association Rules}
\label{sect:association}
The discovery of association rules is a typical problem in data profiling for multiple columns. The algorithm currently used to find association rules needs to scan the database several times. For large data sets, the time overhead of scanning several times is hard to accept. Large amount of data leads to input data, intermediate results and output patterns can be too large to fit into memory and prevents many algorithms from executing \cite{DBLP:journals/mlc/WuFPZY15}. Some scholars have proposed using parallel or distributed methods to solve the problem of data volume \cite{DBLP:journals/tkde/AgrawalS96, DBLP:conf/pdis/CheungHNFF96}. But it is difficult to design parallel or distributed algorithms. 

Therefore, Zaki et al. \cite{DBLP:conf/ride/ZakiPLO97} use sampling to get samples of transaction and find the association rules based on the obtained samples. They take sequential random sampling without replacement as their sampling method and use Chernoff bounds to obtain sample size. Finally, they experimentally prove that sampling can speed up the discovery of association rules by more than an order of magnitude and provide high accuracy for association rules. 

Chen et al. \cite{DBLP:conf/kdd/ChenHS02} propose a two-phased sampling-based algorithm to discover association rules in large databases. At the first stage, a large initial sample of transactions is randomly selected from databases, which is applied to calculate support of each individual item. And these estimated supports are used to trim the initial sample to a smaller final sample S0. At the second stage, association-rule algorithm is performed against the final sample S0 to get association rules according to provided minimum support and confidence. In the experiment, the authors prove 90-95$\%$ accuracy obtained using the final sample S0 and the size of sample is only 15-33$\%$ of the whole databases. This again proves that sampling can be used to speed up data analysis and big data profiling. 

Wu et al. \cite{DBLP:journals/mlc/WuFPZY15} propose an Iterative Sampling based Frequent Itemset Mining method called ISbFIM. The same as \cite{DBLP:conf/ride/ZakiPLO97}, Wu et al. \cite{DBLP:journals/mlc/WuFPZY15} use random sampling as the sampling method. But the difference is that they use iterative sampling to get multiple subsets and find frequent items from these subsets. They can guarantee that the most frequent patterns for the entire data set have been enumerated and implement a Map-Reduce version of ISbFIM to demonstrate its scalability on big data. Because the volume of input data is reduced, the problem that input data, intermediate results, or the final frequent items cannot be loaded into memory is solved. And the traditional exhaustive search-based algorithms like Apriori can be fitted for big data context. 

\subsection{Sampling for Clustering and Anomaly Detection}
\label{sect:cluster}
Clustering is to segment similar records into the same group according to certain characteristics, and those records that cannot be classified into any group may be abnormal points. The challenge that clustering technology encounters in the era of big data is also the problem of data volume, and the clustering operation itself consumes a lot of calculations. Shirkhorshidi et al. \cite{DBLP:conf/iccsa/ShirkhorshidiATH14} divide big data clustering into two categories: single-machine clustering and multiple-machine clustering. Single column reduces the amount of data by using data reduction methods, e.g., sampling and dimensionality reduction. Multi-machine clustering refers to the use of parallel distributed computing frameworks, e.g., MapReduce and cluster resources to increase computing power. 

Kollios et al. \cite{DBLP:journals/tkde/KolliosGKB03} propose biased sampling to speed up clustering and anomaly detection on big data. Unlike the previous work, they consider the data characteristics and analysis goals during the sampling process. Based on the tasks of clustering and anomaly detection, Kollios et al. \cite{DBLP:journals/tkde/KolliosGKB03} consider the data density problem in the dataset. They propose a biased sampling method to improve the accuracy of clustering and anomaly detection. The biased sampling is to make the data points in each cluster and the abnormal points have a higher probability of being selected. In order to achieve this goal, they use the density estimation method to estimate density around the data points. In the experiment, they prove that density-based sampling has a better effect on clustering than uniform sampling. 

\begin{figure}[!t]
	\centering
	\includegraphics[width=3.4in]{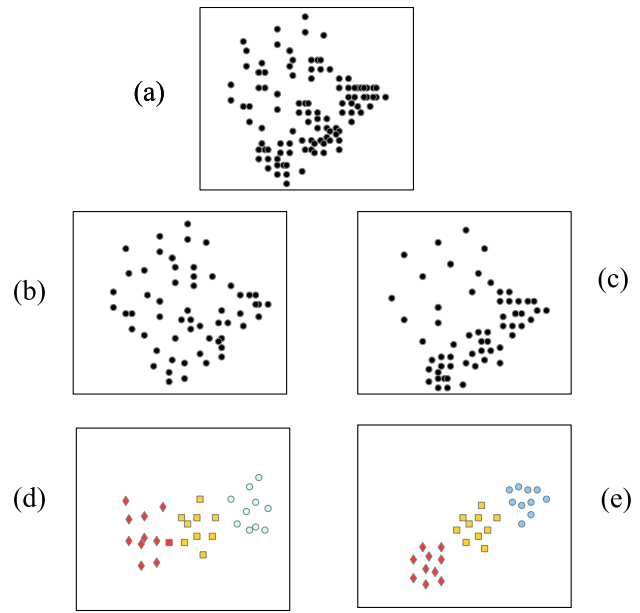}
	\caption{Application of biased sampling in clustering tasks  \cite{DBLP:journals/tkde/KolliosGKB03}.}
	\label{clustering.png}
\end{figure}

Figure \ref{clustering.png} shows the use of biased samples in clustering. Figure \ref{clustering.png}(a) is the distribution of the original data and there are three classes with higher density. Figure \ref{clustering.png}(b) is the result of random sampling on the original data set. Figure \ref{clustering.png}(c) is the result of applying the biased sampling to the original data. Figure \ref{clustering.png}(d) shows 10 data points selected from each of the three categories clustered based on the random sampling, and Figure \ref{clustering.png}(e) shows 10 data points selected from each of the three categories clustered based on the biased sampling method. After comparison with the categories in the original data, it is found that the clustering results of the biased samples are more accurate.

\subsection{Sampling for Summaries and Sketches}
\label{sect:summaries}
Summaries or sketches can be performed by sampling or hashing data values to a smaller domain \cite{DBLP:journals/vldb/AbedjanGN15}. Although different scholars have applied different sampling algorithms, the most commonly used sampling algorithm among data scientists is random sampling \cite{DBLP:conf/ldav/RojasKRD17}. The main reason is that random sampling is the best and easiest to use, which is the only technique commonly used by data scientists to quickly gain insights from big data sets.

Rojas et al. \cite{DBLP:conf/ldav/RojasKRD17}  first interview 22 data scientists working on large data sets and find that they basically use random sampling or pseudo-random sampling. 
Certainly, these data scientists believe that other sampling techniques may achieve better results than random sampling. These scientists perform a data exploration task that used different sampling methods to support classification of more than 2 million generated samples from data records of Wikipedia article edit. Research has shown that sampling techniques other than random sampling can generate insights into the data, which can help focus on the different characteristics of the data without affecting the quality of data exploration and helping people understand the data. This shows that with the application of sampling, Summaries or sketches of data can be created to help scientist observe and understand the data.

Aggregated queries are also a way to generate summaries of data. Aggregate queries are computationally expensive which need to traverse the data. In the era of big data, a single machine often cannot make such a large amount of data. Therefore, aggregate queries for big data are often performed on distributed systems that scales to thousands of machines. The commonly used distributed computing frameworks are Hadoop, spark, etc. Although distributed systems provide tremendous parallelism to improve performance, the processing cost of aggregated queries remains high \cite{DBLP:journals/pvldb/YanCZ14}. Investigation in one cluster of \cite{DBLP:journals/pvldb/YanCZ14} reveals that 90\% of 2,000 data mining jobs are aggregation queries. These queries consume two-thousand machine hours on average, and some of them take up to 10 hours. 

\begin{figure}[!t]
	\centering
	\includegraphics[width=3.4in]{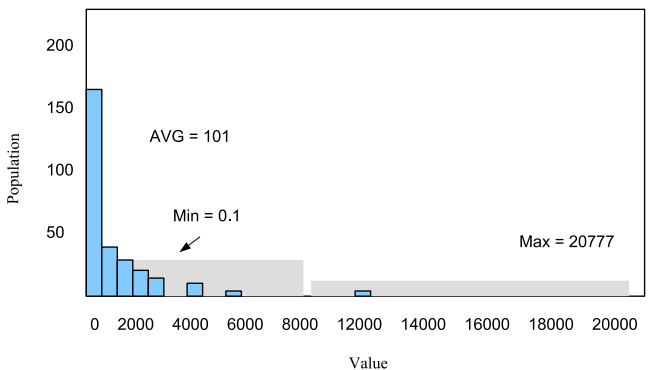}
	\caption{Sparseness of one representative production data \cite{Explained2009Sampling}.}
	\label{SparsenessData.png}
\end{figure}

Therefore, Yan et al. \cite{DBLP:journals/pvldb/YanCZ14} use sampling technique to reduce the amount of data. When error bounds cannot be compromised and data is sparse, they think that conventional uniform sampling often yields high sampling rates and thus deliver limited or no performance gains. For example, uniform sampling with 20\% error bound and 95\% confidence needs to consume 99.91\% of the data whose distribution is shown in Figure \ref{SparsenessData.png}. 
Hence, they propose error-bounded stratified sampling, which is a variant of stratified sampling \cite{Explained2009Sampling} and relies on the insight, i.e., prior knowledge of data distribution, to reduce sample size. Error bound means that the real value has a large probability of falling within an interval. Sparse data means that the data is generally limited but wide-ranging. 

Taking the data distribution in Figure \ref{SparsenessData.png} as an example, error-bounded stratified sampling can divide the data into two groups. One group covers the header data and the other covers the tail data. Because the data range of the first group is small, the sampling rate is also small. Although the data range of the second group is large, the data basically falls in the first group. Even if the data of the second group is all taken as a sample, the overall sampling rate is still low. It is worth mentioning that the technique has been implemented into Microsoft internal search query platform.

\subsection{Sampling for Regression Analysis}
\label{sect:regression}
Statistical analysis such as regression analysis can be used to analyze the relationship between multiple columns in a relation. Sauter \cite{DBLP:journals/technometrics/Sauter05} think that statistics are learned from data. Statistics methods are often used for data profiling, which have encountered the problem of excessive data volume in the era of big data. Statistical analysis of the entire big data set requires a certain amount of calculation and time.

Under the computational pressure of large data sets, many traditional statistical methods are no longer applicable. Although sampling can help with data reduction, how to avoid sampling errors caused by sampling needs to be considered. For example, \cite{wang2019information} mention that in the context of linear regression, traditional sub-sampling methods are prone to introduce sampling errors and affect the covariance matrix of the estimator. Hence, they propose information-based optimal subdata selection method called IBOSS. The goal of IBOSS is to select data points that are informative so that small-sized subdata retains most of the information contained in the complete data. Simulation experiments prove that IBOSS is faster and suitable for distributed parallel computing.

\begin{figure}[!t]
	\centering
	\includegraphics[width=3.4in]{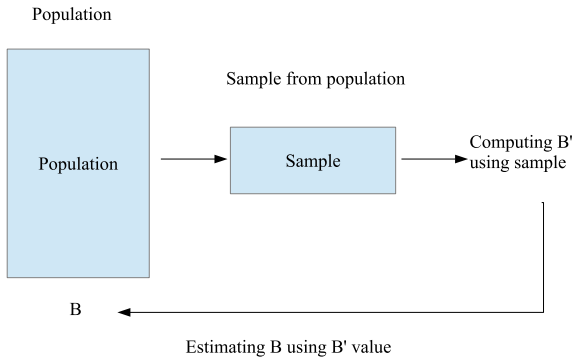}
	\caption{Traditional big data regression analysis \cite{jun2015divided}.}
	\label{traditional_big_data_analysis.png}
\end{figure}

\begin{figure}[!t]
	\centering
	\includegraphics[width=3.4in]{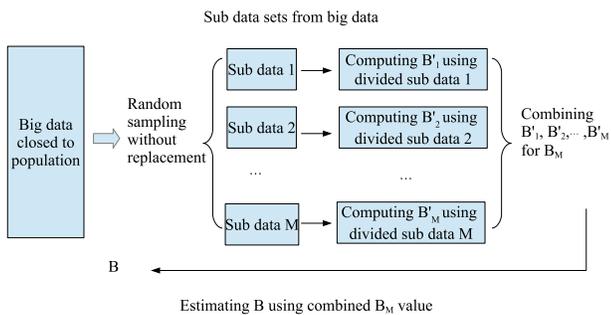}
	\caption{Sampling-based partitioning big data regression analysis \cite{jun2015divided}.}
	\label{sampling_to_divided_big_data.png}
\end{figure}

Jun et al. \cite{jun2015divided} propose to use sampling to divide big data into some sub data sets in regression problem for reducing the computing burden. The traditional statistical analysis of big data is to sample from big data, and then perform statistical analysis on the sample to infer the population. Jun et al. \cite{jun2015divided} divide the big data closed to population into some sub data sets with small size closed to sample which is proper for big data analysis. They treat the entire data set as a population and the sub set as a sample to reduce computing burden. And they select regression analysis to perform experiments. The traditional processing is shown in Figure \ref{traditional_big_data_analysis.png}, and their design is shown in Figure \ref{sampling_to_divided_big_data.png}. 

Their design consists of three steps: the first step is to first generate M sub-data sets using random samples without replacement; the second step is to calculate the regression parameters of each sub-data set and calculate the average of regression parameters of the M sub-data sets; the third step is to use the averaged parameters obtained in the second step to estimate regression parameters on the entire data set. This design that combines sampling and parallel processing helps them speed up regression analysis on big data. By experimenting with the data set from the simulation and UCI machine learning repository, the author proves that the regression parameters obtained by distributed calculation on random samples are close to the regression parameters calculated on entire data set. This provides a reference for statistical analysis on the entire large data set.

\section{Sampling for Dependencies} \label{dependencies}

A dependency is a metadata that describes the relationship between columns in relation,
based on either value equality or similarity \cite{DBLP:journals/tods/Song011}. 
There are many use cases for dependencies. 
For example, unique column combinations are used for finding key attributes in relation \cite{DBLP:conf/vldb/SismanisBHR06}, and functional dependencies can be used for schema normalization \cite{DBLP:conf/edbt/PapenbrockN17} or consistent query answering \cite{DBLP:conf/sigmod/LianCS10}, 
while inclusion dependencies can suggest how to join two relations \cite{DBLP:journals/vldb/AbedjanGN15}.
Inclusion dependencies together with functional dependencies form the most important data dependencies used in practice \cite{DBLP:journals/is/LopesPT02}. But discovery of dependencies is time consuming and memory consuming. Many functional dependencies discovery algorithms are not suitable for large data sets. 
Sampling could be employed to estimate the support and confidence measures of data dependencies \cite{DBLP:conf/icde/SongCC12,DBLP:journals/tkde/Song0C14}.
By sampling, you can select a small enough representative data set from the big data set. Hence, the choice of sampling method is very important, which help to ensure that the estimated inaccuracy rate is below a predefined bound with high confidence. 
Specifically, based on the classification for dependency in \cite{DBLP:journals/vldb/AbedjanGN15}, we investigate sampling for unique column combinations in Section \ref{sect:unique}, functional dependency in Section \ref{functional dependencies}, inclusion dependency in Section \ref{sect:inclusion}.

\subsection{Sampling for Discovery of Unique Column Combinations}
\label{sect:unique}
An important goal in data profiling is to find the right key for the relational table, e.g., primary key. The step before key discovery is to discover unique column combinations. Unique column combinations are sets of columns whose values uniquely identify rows, which is an important data profiling task \cite{DBLP:journals/pvldb/HeiseQAJN13}. But discovery of unique column combinations is computationally expensive, which is suitable for small dataset or samples of large dataset. For large data set, sampling is a promising method for knowledge discovery \cite{DBLP:conf/pods/KivinenM94}. Based on sampling-based knowledge discovery, it is necessary to first select samples from the entire data set and obtain knowledge from the samples, and then use the entire data set to verify that the acquired knowledge is correct.

A typical algorithm for identifying key attributes is GORDIAN proposed by Sismanis et al. \cite{DBLP:conf/vldb/SismanisBHR06}. The main idea of GORDIAN is to turn the problem of keys identification into cube computation problem, and then find non-keys through cube computation. Finally, GORDIAN calculates the complement of the non-keys set to obtain the desired set of keys. Therefore, the GORDIAN algorithm can be divided into three steps: (\textit{i}) create the prefix tree through a single pass over the data; (\textit{ii}) find maximal non-uniques by traversing the prefix tree with prunning; (\textit{iii}) get minimal keys from set of maximal non-uniques. In order to make GORDIAN scalable to large datasets, Sismanis et al. combine GORDIAN with sampling. Experiments have shown that sampling-based GORDIAN can find all true keys and approximate keys using only a relatively small number of samples.

GORDIAN algorithm is further developed by Abedjan and Naumann \cite{DBLP:conf/cikm/AbedjanN11} to discover unique column combinations. Since the existing algorithms are either too violent or have high memory requirements and cannot be applied to big data sets. A hybrid solution HCA-Gordian, which combines Gordian algorithm \cite{DBLP:conf/vldb/SismanisBHR06} and their new algorithm the Histogram-Count-based Apriori Algorithm (HCA), is proposed by Abedjan and Naumann \cite{DBLP:conf/cikm/AbedjanN11} to discover unique column combinations. GORDIAN algorithm is used to find composite keys and the HCA is an optimized bottom-up algorithm which takes efficient candidate generation and statistics-based pruning methods. HCA-Gordian performs Gordian algorithm on a smaller sample of table to discover non-uniques and non-uniques will be used as pruning candidates when executing HCA on the entire table. 

In the experiment setup, the sample size for the preprocessing step of non-unique discover is always 10,000 tuple sample. Especially when the amount of data is large and the number of unique is small, the runtime of HCA-Gordian is lower than Gordian. For example, when using real world tables for experiments, search speed of HCA-Gordian is four times faster than Gordian. And as the data set grows larger, e.g., the National file contains 1,394,725 tuples, Gordian takes too long to run, while HCA-Gordian only takes 115 seconds to complete. In addition, When the number of detected non-uniques is high, the discovery effect of HCA-Gordian is better than Gordian.

\subsection{Sampling for Functional Dependencies}
\label{functional dependencies}
A functional dependency refers to a set of attributes in a relationship that determines another set of attributes. For example, there is such a functional dependency $A$->$B$, which means that any two records in the relationship, when their values on the attribute set $A$ are equal, the values on the attribute set $B$ must be equal. Bleifuss et al. \cite{DBLP:conf/cikm/BleifussBFRW0PN16} propose an approximate discovery strategy AID-FD (Approximate Iterative Discovery of FDs) which sacrifices a certain correct rate in exchange for performance improvement. AID-FD uses an incremental, focused sampling of tuple pairs to deduce non-FDs until user-configured termination criterion is met. The authors have demonstrated in experiments that the AID-FD method uses only 2\%-40\% of the time of the exact algorithm when processing the same data set, but finds more than 99\% of the functional dependencies.

Papenbrock and Naumann \cite{DBLP:conf/sigmod/PapenbrockN16} mention that today's various functional dependencies discovery algorithms do not have the ability to process more than 50 columns and 1 million rows of data.
Thus, they propose the sampling-based FD discovery algorithm HYFD.
And there are three properties in sampling-based FD discovery algorithms: Completeness, Minimality, Proximity, which are important for HYFD. HYFD combines column-efficient FD induction techniques with row-efficient FD search techniques in two phases. In Phase 1, they apply focused sampling techniques to select samples with a possibly large impact on the result’s precision and produce a set of FD candidates based on samples. In Phase 2, the algorithm applies row-efficient FD search techniques to validate the FD candidates produced in Phase 1. The sampling method allows functional dependencies discovery algorithms to be extended to large data sets. 

In experiments, when the data set is not very large, the runtime of HYFD is almost all lower than other algorithms. When the data set exceeds 50 columns and 10 million rows, HYFD can get the result through a few days of calculation. However, other algorithms cannot complete the calculation, because the time complexity for these algorithms is exponential. This again demonstrates that sampling is important for data profiling, e.g., FD discovery.

In the above, we mention that using focused sampling to find functional dependencies.  In this section, we will mention the use of random sampling to find soft functional dependencies. The so-called "soft" functional dependency is relative to the "hard" functional dependency. A "hard" functional dependency means that the entire relationship satisfies the functional dependency, while a "soft" functional dependency means that the entire relationship is almost satisfied, or that there is a high probability of satisfying the functional dependency. 

Ilyas et al. \cite{DBLP:conf/sigmod/IlyasMHBA04} propose sampling-based CORDS, which means that automatic discovery of correlations and soft functional dependencies between columns, to find approximate dependencies. Among them, correlation refers to the general statistical dependence, while soft functional dependence refers to that value of attribute C1 determines the value of attribute C2 with high probability. CORDS use enumeration to generate pairs of columns that may be associated, and heuristically cuts out those unrelated column pairs with high probable. CORDS apply random sampling with replacement to generate sample. In the implementation of CORDS, they only use a few hundred rows of sample data, and the sample size is independent of the data size. In the experiment to evaluate the advantages of applying CORDS, where run a workload of 300 queries on the Accidents database, the median query execution time and worst query execution time with CORDS applied were better than those without CORDS. 
Hence, CORDS is efficient and scalable when it encounters large-scale dataset. 

Approximate functional dependence is similar to the meaning of soft functional dependency. Approximate functional dependence requires the normal functional dependency to be satisfied by most tuples of relation $R$ \cite{DBLP:conf/icdt/KivinenM92, DBLP:journals/tkde/LiuLLC12}. Of course, approximation functional dependencies contain exact functional dependencies that are satisfied throughout the relationship. As mentioned in \cite{DBLP:conf/icdt/KivinenM92}, when the amount of data is large, the time for discovery of functional dependency will increase exponentially.  Therefore, Kivinen and Mannila \cite{DBLP:conf/icdt/KivinenM92} propose to discover approximate dependencies by random sampling. In fact, sampling can be used not only to find approximate functional dependencies, but also to verify exact functional dependencies \cite{DBLP:journals/tkde/LiuLLC12}. If the exact functional dependency does not satisfy all the sample data, then the whole relationship is definitely not satisfied, hence such functional dependencies can be removed.

Functional dependencies are  satisfied for all tuples in the relation, while conditional functional dependencies (CFDs) is to hold on the subset of tuples that satisfies some patterns \cite{DBLP:conf/icde/BohannonFGJK07}. And CFDs can be used for data cleaning \cite{DBLP:conf/icde/BohannonFGJK07, DBLP:journals/tods/FanGJK08}. Fan et al. \cite{DBLP:journals/tkde/FanGLX11} propose three methods for conditional functional dependencies discovery. However, when the size of data set is large, no dependency discovery algorithms scale very well to discover minimal conditional functional dependencies.

When mining CFDs on big data, the volume issue of big data has to be solved. Li et al. \cite{8935096} develop the sampling algorithms to obtain a small representative training set from  large and low-quality datasets and discover CFDs on the samples. They use sampling technology for two reasons. One is that finding CFD needs to scan the data set multiple times, and sampling helps reduce the amount of data. The second is to use the sampling method to help them filter those dirty items on the low-quality data set and choose clean items as the training set. They define criteria for misleading tuples, which are dirty, incomplete or very similar to popular tuples. And then they design a Representative and Random Sampling for CFDs (BRRSC), which is similar to reservoir sampling \cite{DBLP:journals/toms/Vitter85}. The difference is that they combine the criteria defined above during the sampling process. Furthermore, they propose fault-tolerant CFDs discovery and conflict-resolution algorithms to find CFDs. Finally, experimental results show that their sampling-based CFD discovery algorithms can find valid CFD rules for billions of data in a reasonable time.

\subsection{Sampling-based Test for Inclusion Dependency Candidates}
\label{sect:inclusion}
The definition of inclusion dependencies (INDs) is that the combination of values that appear in a set of attribute columns must also appear in another set of attribute columns \cite{DBLP:conf/edbt/MarchiLP02}. Therefore, inclusion dependencies are often used to discover foreign keys \cite{DBLP:journals/is/LopesPT02}. However, discovery of inclusion dependencies is computationally expensive. One of the reasons is that the existing algorithms need to shuffle huge amounts of data to test inclusion dependencies candidates, which puts pressure on both computing and memory \cite{DBLP:conf/btw/0001PDFHZZN17}. 

Under these circumstances, Kruse et al. \cite{DBLP:conf/btw/0001PDFHZZN17} propose fast approximate discovery of inclusion dependencies (FAIDA). FAIDA can guarantee to find all INDs and only false positives with a low probability in order to balance efficiency and correctness. FAIDA uses algorithms \cite{DBLP:journals/jiis/MarchiLP09, DBLP:journals/pvldb/PapenbrockKQN15} of Apriori-style to generate inclusion dependencies candidates. The inverted index values and operates on a small sample of the input data. The sampling algorithm is applied to each table to get each sample. Rather than use random sampling to get sample, they assure that sample table contains min \{s, d$_{A}$\} distinct values for each column $A$, where s represents sample size and d$_{A}$ represents number of distinct values in column $A$. 

In their experiments, they set sample size to a default of 500. In order to verify the efficiency of FAIDA, Kruse et al. \cite{DBLP:conf/btw/0001PDFHZZN17} compare FAIDA's runtime with the state-of-the-art algorithm for exact IND discovery BINDER \cite{DBLP:journals/pvldb/PapenbrockKQN15} on multiple datasets. On four datasets, FAIDA is steadily 5 to 6 times faster than BINDER, and they generate and test almost the same number of IND candidates. Especially when one of the datasets reaches 79.4GB, BINDER takes 9 hours and 32 minutes to complete, while FAIDA only takes 1 hour and 47 minutes.
Their evaluation shows that sampling-based FAIDA outperforms the state-of-the-art algorithm by a factor of up to six in terms of runtime without reporting any false positives.

\section{Summary and Future Works}
\label{summary}
Data in various fields are increasing on a large scale. Big data brings us new opportunities and challenges. Through data analysis and data mining of big data, we can get a lot of potential value. However, due to the large amount of data, it brings great challenges to the processing and storage. Therefore, data analysis, data mining or data profiling on large data sets have to face the pressure of calculation and time. Increasing computing power by using clusters of computers is one solution, but many times this is not the case, and designing distributed computing is often difficult. 
Hence, the application of data reduction techniques like sampling is very important. There are some mature research articles on data profiling and sampling, 
while little attention is paid to sampling and profiling over big data, therefore this article focuses on researching sampling and profiling in big data context. 
We first give a brief introduction of data profiling and introduce some important factors of sampling in detail. 
Then, according to the classification of data profiling in \cite{DBLP:journals/vldb/AbedjanGN15}, we introduce the application of sampling in single column data profiling, multiple columns data profiling and dependency discovery. In conclusion, Table \ref{table:summary-of-sampling} summarizes the sampling for data profiling tasks investigated in survey, indicating the widespread use of sampling in data profiling.

The above survey on ``sampling and profiling over big data'' is mainly about relational databases, and rarely involves graph data or time series data. Since there is less research on sampling-based data profiling for graph data or time series data, we provide some future directions as follows.

\subsection{Sampling for Profiling Time Series Data}

Many tasks on time series data need data profiling, e.g.,  
matching heterogeneous events in a sequence \cite{DBLP:conf/sigmod/ZhuSL0Z14,DBLP:journals/tkde/GaoSZWLZ18} with profiled patterns \cite{DBLP:conf/icde/ZhuSWYS14,DBLP:journals/tkde/SongGWZWY17},
cleaning time series data \cite{DBLP:journals/pvldb/ZhangS0Y17} under speed constraints \cite{DBLP:conf/sigmod/SongZWY15}, or
repairing timestamps according to the given temporal constraints \cite{DBLP:journals/pvldb/SongC016}
such as sequential dependencies \cite{DBLP:journals/pvldb/GolabKKSS09}. 
All these studies use data profiling to detect and repair erroneous temporal data. 
The computational cost and time cost in large-scale temporal data streams can be high. Therefore, sampling for profiling time series data is valuable and necessary.

In the time series data stream, we do not need to get exact results, e.g., when calculating the quantiles or probability distributions of speeds. Approximate results are valuable in time-series data streams, for example approximate probability distributions of speeds can also help us perform effective anomaly detection. In the sampling of time series data, statistical probability distributions of speeds are different from discovering quantiles. The speed of time series data depends on the adjacent time-series data points, which means that sampling for calculating speed of time series requires a set of data points in a window. Therefore, how to apply the sampling technology to the aforesaid data profiling task of time series data needs further experimental analysis and research.

\subsection{Sampling for Profiling Graph Data}
Data profiling is also heavily used in graph data, e.g., 
using Petri Nets in process mining to recover missing events \cite{DBLP:journals/pvldb/0001SZL13,DBLP:journals/tkde/0001SZLS16} and clean event data \cite{DBLP:conf/icde/0001SLZP15}, 
discovering keys for graphs and applying keys to study entity matching \cite{DBLP:journals/pvldb/FanFTD15},
or
defining functional dependencies for graphs \cite{DBLP:conf/sigmod/FanWX16a} and discovering them \cite{DBLP:conf/sigmod/FanHLL18}.
However, the above studies still seem to be difficult when encountering large graphs. 
Fan et al. \cite{DBLP:journals/pvldb/FanFTD15} prove that entity matching is NP-complete for graphs and recursively defined keys for graphs bring more challenges. 
In this case, one has to design two parallel scalable algorithms, in MapReduce and a vertex-centric asynchronous model. In order to find Graph Functional Dependencies, Fan et al. \cite{DBLP:conf/sigmod/FanHLL18} have to deal with large-scale graphs by designing effective pruning strategies, using parallel algorithms, and adding processors. As mentioned earlier, designing parallel algorithms is difficult. 

Equivalently, profiling for graph data has to face the pressure of computing and memory when data profiling encounters large graphs. Therefore, it is necessary and worth researching to sample the graph data and carry out the tasks of data profiling based on the sample. But sampling graph data is more difficult than sampling relational data. Leskovec and Faloutsos \cite{DBLP:conf/kdd/LeskovecF06} did practical experiments on sampling from large graphs. They concluded that best performing methods are the ones based on random-walks and "forest fire", with sample sizes as low as 15$\%$ of the original graph. However, how to apply these graph sampling methods to the above-mentioned graph data-based data profiling tasks is waiting for further experiments and exploration.

\subsection{Sampling for Profiling Heterogeneous Data}
Data profiling is also widely used for heterogeneous data, e.g., 
discovering matching dependencies (MDs) \cite{DBLP:conf/cikm/SongC09,DBLP:journals/dke/Song013}, 
reasoning about matching rules \cite{DBLP:journals/pvldb/FanJLM09,DBLP:journals/vldb/FanGJLM11}, 
discovering a concise set of matching keys \cite{DBLP:journals/pvldb/Song0C14} 
and conditional matching dependencies (CMDs) \cite{DBLP:journals/tkdd/WangSCYC17}.
However, these profiling tasks also have to face computational pressure in a big data context. 

In fact, MDs, DDs and data dependencies are all based on differential functions. 
When calculating the measures for differential dependencies, 
performing sampling of pairwise comparison is more difficult. 
Given an instance of relation R with N data tuples, pairwise comparison M will increase the total number to $\frac{N*(N-1)}{2}$, which will greatly increase the number of populations. 
However, many pairs in M are meaningless when calculating support for DDs \cite{DBLP:conf/adc/KwashieLLY15}, 
which means that the proportion of pairs we want is very small. Therefore, we must increase the sampling rate to expect to include these pairs in the sample, so as to get the approximate results as close as possible.

%

\bibliographystyle{unsrt}
\bibliography{ref.bib}
%

\end{document}